# Is the Consequence of Superluminal Signalling to Physics Absolute Motion through an Ether?

Remi Cornwall
Future Energy Research Group
Queen Mary, University of London, Mile End Road, London E1 4NS

## Abstract

In an earlier paper the author expounded an interferometer scheme to communicate classical data over an entangled quantum channel. We return to this concept to show that the laws of Quantum Mechanics are not violated and that the device is able to affect the statistical blend of the quantum states (and that this can be detected) but not the statistics – i.e. the physics of observables. The ideas of superluminal information transfer, discussed in the previous paper, are taken forward to develop a notion of absolute space, time and motion with relativistic effects ascribed to motion through an absolute reference frame – as the logical consequences dictate, permeated with a material causing the 'relativistic' effects. The reciprocal nature of the Lorentz Transform is shown to fail under superluminal signalling – one frame will be *absolutely* time dilated and length contacted; thus a full 'Ether Transformation' (though this cannot be a group) and a velocity addition law are derived, the Twin's Paradox is reconsidered. It would seem, in Special Relativity at least, that the phenomenological effects of motion are placed in an absolute, logical, materialistic setting, rather than a confusing relative one that, perhaps, allows no further inspection of just 'what it is' causing the effects. The author then discusses whether this programme can be carried forward into General Relativity with the space-time distortions ascribable to changes in the properties of the ether but on the whole describable in 3-space.

## 1. Introduction

An interferometer set-up[1] utilising entangled particles to receive information, by remote change in quantum state, by a modulator, has been criticised as not obeying the laws of Quantum Mechanics in that local changes in quantum state cannot affect remote physics[2-4]. Naturally this Bell Channel setup[5, 6] alludes to the EPR paradox[7, 8].

The communication scheme essentially prepared particles in nearly pure, entangled states with one stream headed to the modulator and the second stream headed to the interferometer. The act of measurement and wave-function collapse[9-12] turns the practically pure states into a statistical mix[13] of orthogonal states which the interferometer and rotation filters is able to converge and interfere such that pure and mixed blends can be ascertained. A protocol is then established to send classical data over this quantum channel such that the act of measurement (and hence a statistical blend) would signify one bit and no measurement, the complement bit. The next section shall discuss why this scheme doesn't conflict with the laws of Quantum Mechanics.

This paper then recounts and expands on the absolute space and time notions[i] developed in the earlier paper[1]. Superluminal signalling[ii] is shown to break the reciprocity implicit in the Interval[11, 14, 15] and the Lorentz Group by placing the Doppler effect in the light of SL communication. Space-time diagrams viewed from the rest frame show the situation that one frame must, absolutely, be time dilated and length contracted. Then systematically a transform is developed, in flat space at least, from the Universal Rest Frame to a moving frame and a velocity addition law. Space-time appears to be deconstructed with time separate from spatial variables.

The Twin's Paradox is analysed from the universal rest frame giving a logical basis for this counter-intuitive scenario – the only conclusion is that particles travel (and accelerate) through a medium which effects their rate of time and dimension compared to the rest frame. This medium or "Ether" was meant to be dismissed by Special Relativity but was ironically brought back by the quantum theory of fields[9, 16] – just how can an "empty vacuum" manifest particles? Sakharov[15iii] proposed that gravity was the "metric elasticity" of space associated with quantum fluctuations of other fields. Indeed the ether is imbued with mechanical properties as Maxwell originally believed[17].

[i] 'Relativity' is just the metrology of light-speed limited signals with time dilation and length contraction effects.

[ii] Which by transmission of pure information by quantum state has no mass-energy and doesn't violate Relativity

[iii] pp. 426-428

The author then asks if the programme of research can be extended to General Relativity. The vacuum, far away from gravitating bodies ("free-space") offers a gold standard in absoluteness but one must contend with metrics that are time dependent, frame dragging or even the weirder, perhaps science-fiction concepts of wormholes and discontinuous space or time-travel. An infinitesimal section of space in GR is flat and the result of decoupling time from spatial variables in this paper, by the ether transform, may well limit the more fantastical solutions in GR, such as, discontinuous space or time-travel.

SL signalling allows a *time-standard* to be communicated from far, from free-space such that clocks can be synchronised even in non-static metrics with a global world-time (it is already an old result that in a gravitational well time *is* already running *absolutely* slower). This SL communication of a time standard allows the communication of *distance-standard* by sending a time domain equivalent of a unit length of invariant c-speed travel; once the effect of time dilation is removed, a direct comparison between the true distance standard and the local can be carried out. A network of such comparison stations throughout the gravitating region communicating their results *superluminally*, back to the mapping station in free-space, allows a *near instant snapshot* of the region and each station to be located in flat 3-space with a value of length and time dilation *at that location* to be worked out.

If the programme can be completed by the Ether transform (section 5) decoupling time from space co-ordinates, then curved space-time is replaced with flat Euclidian 3-space, permeated with a material, an ether, through which motion gives "Relativistic" effects and whose perturbation by mass-energy gives gravity.

## 2. No conflict with the Laws of Quantum Mechanics

Arguments have been put forward for a number of years that SL signalling is impossible in the framework of Quantum Mechanics[2-4]. A quick summary of this is: a local trace (measurement) performed on the total wave-function for the system (tensor product) has no effect on the statistics (the physics) of the remote system.

Quantum Mechanics is founded on a limited number of principles. Relevant here in discussing the impossibility proofs is the measurement principle. The expected value of an operator, A is:

$$\langle A \rangle = \langle \psi | A | \psi \rangle \text{ or } \langle A \rangle = tr[A\rho] \quad \text{eqn. 1}$$

This is a statement of the physics of observables: such operators are Hermitian[9-11] and so equal to their own complex transpose. Naturally such a matrix is symmetrical and the diagonal is real valued. The trace is precisely the operation of summing the diagonal. For instance a state $|\psi\rangle$ or a state with a phase shift $e^{i\theta}|\psi\rangle$ or a blend with the same amplitude, that mixes states coherently or incoherently will give the same result.

It is quite clear that the interferometer setup for the communication device[1] is able to measure statistical blends – that is, it can tell the difference between practically pure and mixed states. One bit is represented as a practically pure state and is thus capable of interference, whilst the other bit is (after wave-function collapse by the modulator) is a mixed state and cannot interfere.

| Measurement/Modulation at distant system and state of two photon system | State of distant system | State of local system | Local measurement by interferometer after modulation of distant system |
|---|---|---|---|
| No modulation: 'Binary 0' $\frac{1}{\sqrt{2}}\left(\lvert H\rangle_1 \lvert V\rangle_2 + \lvert H\rangle_2 \lvert V\rangle_1\right)$ | Entangled => Pure state $\frac{1}{\sqrt{2}}\left(\lvert H\rangle_1 + \lvert V\rangle_1\right)$ (Or at least some superposition) | Entangled => Pure state $\frac{1}{\sqrt{2}}\left(\lvert V\rangle_2 + \lvert H\rangle_2\right)$ | Pure state results in interference (Or at least some interference since source is not ideally pure) |
| Modulation: 'Binary 1' $\frac{\lvert H\rangle_1 \lvert V\rangle_2}{\sqrt{2}}$ *or* $\frac{\lvert H\rangle_2 \lvert V\rangle_1}{\sqrt{2}}$ | Not entangled <=> Mixed state $\frac{\lvert H\rangle_1}{\sqrt{2}}$ *or* $\frac{\lvert V\rangle_1}{\sqrt{2}}$ | Not entangled <=> Mixed state $\frac{\lvert H\rangle_2}{\sqrt{2}}$ *or* $\frac{\lvert V\rangle_2}{\sqrt{2}}$ | Mixed state gives no interference |

Figure 1 – The Protocol

This hasn't affected the physics at all: a photon detector without the interferometer will still measure (Appendix 1) by eqn. 1:

$$\frac{1}{\sqrt{2}}\left(|H\rangle + |V\rangle\right) \text{ or } \frac{1}{2}|H\rangle\langle H| + \frac{1}{2}|V\rangle\langle V|$$

## 3. Relativism vs. Absolutism

It was obvious from its inception that the Maxwell equations of the electromagnetic field weren't subject to Galilean transformation:

$$\left(\nabla^2 - \frac{1}{c^2}\frac{\partial^2}{\partial t^2}\right)\psi = 0 \qquad \text{eqn. 2}$$

Lorentz suggested a transformation to affect changes between different reference frames of electromagnetic signals. The very real electric and magnetic fields (in the sense of their measurement, energy and momentum[17]) suggested that light travelled through a medium with mechanical properties - The Ether, much as sound through air.

It would seem necessary after the Lorentz transform, that if the transform applied to electromagnetic effects, affecting length and time measurements, then it should affect other areas of physics too and the effects ("Lorentz contractions", "Fitzgerald dilation") would be called "Ether drag". However the Michelson-Morley experiment was *null* and failed to find any relative motion to the electromagnetic ether.

No faster signal was known to physics at the time and Einstein began to view the Lorentz transformation and the interval as fundamental to space and time instead of an ether theory. The invariant interval can be derived straight from electromagnetic waves and is the cornerstone of his theory:

$$\left(\nabla_A^2 - \frac{1}{c^2}\frac{\partial^2}{\partial t_A^2}\right)\psi_A = 0 = \left(\nabla_B^2 - \frac{1}{c^2}\frac{\partial^2}{\partial t_B^2}\right)\psi_B$$

Substitute the general solution: $\psi = e^{\mathbf{k}\cdot\mathbf{r}-\omega t}$

Which results in: $c^2T^2 - \lambda^2 = 0$

Where T is the period and $\lambda$ is the wavelength

And the interval is found:

$$c^2t_A^2 - x_A^2 - y_A^2 - z_A^2 = c^2t_B^2 - x_B^2 - y_B^2 - z_B^2 \quad \text{eqn. 3}$$

Relativism was the new Absolute. This is a Logical Positivist[18, 19] approach that has characterised much of 20$^{th}$ Century physics, to the despair of some[iv].

Mental models worked well in the 18$^{th}$ and 19$^{th}$ centuries, although they can be just plain wrong but inspiring of progress too. Models and intuition are much under-valued today. Einstein was a man with great intuitive gifts and he rejected much of Quantum Mechanics (including non-locality) and it is ironic that today it should come back to break a model-less, even counter-intuitive worldview that Relativity has given us.

How so? The symmetry and reciprocity of the Lorentz group has seduced us with beautiful mathematics and we are soon taught to forget mind bending paradoxes defying logic and concepts of space, time, quantity and order:

- If two events are exactly simultaneous in one frame, how are they perceived as not in another frame?

- If frame A is time dilated/length contracted relative to frame B, how can frame B be too, relative to A?

- If a pair of twins is each in an inertial frame (one twin going on a journey coasting most of the way) how is it that one ages more than the other yet time dilation is meant to be reciprocal?

- If frame A is head on to frame B and a third frame says they are both moving head-on or receding from each other at 'c', how can they only measure relative velocity as 'c' and not '2c'?

- That we must give up on universal time and universal distance measurement in GR. The Universe's constituent matter exists then, where and when?

The earlier paper[1] answered the first question by referring back to a Universal Rest Frame by SL signals which rendered events simultaneous in one frame to be simultaneous in all. This covered both failure of simultaneity in time and failure of simultaneity at a distance. Now the principle is expanded for the other situations in this paper, to show how SL signalling:

- Breaks the *reciprocity* of the interval/Lorentz transform – so that one frame *is* time dilated/length contracted. Shows how these paradoxes are due to the finite speed of light *and* the relativistic dilations, giving a null, reciprocal effect on some occasions (Doppler shift) and asymmetry on others (Twin's Paradox, section 6).

- Derivation of an Ether Transform (section 5) that can refer all measurements back to a Rest Frame in a systematic manner.

- Breaks the separation/closure velocity paradox (section 5.1.1).

- Allows universal clock synchronisation and universal distance measurement over all space.

---

[iv] In other words: give up real insight and work in terms of only what you can measure and model mathematically.

4. Breaking the reciprocity of time dilation

Appendix 2 gives an account of a classic Twin's Paradox problem. It is a good place to start when constructing the notion of an ether since the effect occurs when one twin moves away from an agreed rest frame. The invariant interval (eqn. 3) between the two frames is the conventional explanation.

A space-time diagram approach (appendix), though only from the home's perspective, plots the yearly beacons from the home base and the moving twin. It finds the reason for the asymmetry in the number of beacons counted due to a combination of time dilation (Lorentz transform) and a lengthening/foreshortening effect from "running away"/"chasing down" of beacons (a Doppler effect) on the outward and return legs.

Using this space-time diagram we now consider the fundamental aspect of this problem to be the Doppler shift, as the twin paradox is essentially two legs of Doppler shift. Appendix 3 shows the home view space-time diagrams for the beacons at home and the beacons from the moving twin. The analysis assumes no specific form for the time dilation but just calls it Γ(v). The known result that the Doppler shift is reciprocal fixes this function as γ(v), the time dilation factor and the Doppler shift formula is obtained. After the next sub-section and section 5 we will return to the Twin's Paradox.

4.1.1. Use of Superluminal Signals

The situation with SL signals for the same home perspective view of beacons being sent and received is now changed. Being superluminal, the signals are then *horizontal* on the diagram. Appendix 3's analysis with light-speed signals lead to world lines for the signals sloped at an angle of 45°. If 'c' tends to infinity both expressions for the Doppler shift, from both frames, take on the same form and the relative velocity must only enter via expressions for the time dilation: $\Gamma_A(v)$ and $\Gamma_B(v)$.

A logical conflict occurs if the time dilations are to be reciprocal for all v:

- If v is non-zero then to maintain reciprocity, $\Gamma_A(v)$ and $\Gamma_B(v)$ must be independent of v and time dilation wouldn't occur (B could do a Twin's paradox round trip and we that know it does).
- OR both frames started moving such that $\Gamma_A\left(\mp\frac{v}{2}\right) = \left|\Gamma_B\left(\pm\frac{v}{2}\right)\right|$ which is an even function but one frame never applied a force to the other, so this cannot be so.

This means for SL signals that the velocities can't be relative between the frames. Using a known result: consider, then, frame A to be at absolute rest and frame B is sending light-speed signals that are Doppler shifted, so $\Gamma_B(v) = \gamma$, as was proven in the previous section. The velocity would then be absolute and this would also apply to the case of sending SL signals too.

The conclusion is: when sending superluminal signals to measure time dilation between frames, each frame is time dilated by its absolute velocity $\gamma(v_A)$ or $\gamma(v_B)$. Gamma increases monotonically so SL signalling of a time standard between frames is definitely not reciprocal.

5. The Ether Transform

The earlier paper presented a 1D Ether transform and we generalise it here. Note that bodies with mass-energy still cannot move faster than 'c'. Relativistic effects occur parallel to the velocity vector and the time delay terms are dropped due to SL signalling[1]:

$$T = \gamma\left(t - \frac{r\cdot v}{c^2}\right) \mapsto T = \gamma t$$
$$R = r_\perp + \gamma\left(r_\parallel - vt\right) \mapsto R = r_\perp + \gamma r_\parallel \qquad \text{eqn. 4}$$

Where $\gamma = \left(1 - v^2/c^2\right)^{-1/2}$, the Rest Frame is made special with capitalisation of T and R.

Since $r = r_\perp + r_\parallel$ and $r_\parallel = \frac{(r\cdot v)v}{v^2}$ we can write the transformation in the spatial vector as:

$$R = \left(r - r_\parallel\right) + \gamma r_\parallel$$
$$\Rightarrow R = r + (\gamma - 1) r_\parallel$$

And so:

$$\Rightarrow R = r + \frac{(\gamma-1)}{v^2}(r\cdot v)v \qquad \text{eqn. 5}$$

In matrix form the Ether transformation is:

$$\begin{bmatrix} T \\ X \\ Y \\ Z \end{bmatrix} = \begin{bmatrix} \gamma & 0 & 0 & 0 \\ 0 & 1+(\gamma-1)\frac{v_x^2}{v^2} & (\gamma-1)\frac{v_x v_y}{v^2} & (\gamma-1)\frac{v_x v_z}{v^2} \\ 0 & (\gamma-1)\frac{v_x v_y}{v^2} & 1+(\gamma-1)\frac{v_y^2}{v^2} & (\gamma-1)\frac{v_y v_z}{v^2} \\ 0 & (\gamma-1)\frac{v_x v_z}{v^2} & (\gamma-1)\frac{v_y v_z}{v^2} & 1+(\gamma-1)\frac{v_z^2}{v^2} \end{bmatrix} \begin{bmatrix} t \\ x \\ y \\ z \end{bmatrix} \qquad \text{eqn. 6}$$

The spatial matrix can be made diagonal but the resulting transformation would change the orientation of the vector [t x y z] to the absolute vector [T X Y Z].

The notion of space-time is deconstructed and the time 'co-ordinate' is once again special. What follows for SR will follow too for GR in the limit of infinitesimal regions obeying SR. The reciprocity and symmetry of the Lorentz transform is lost and so the transformation doesn't form a group, thus a velocity addition law cannot be derived in a straight forward manner.

## 5.1. Addition of Velocities

Once again, relativistic effects occur parallel to the velocity vector so we start by considering 1D velocity addition and subtraction. It is clear in the case of addition of velocities, that the resultant time dilation will be a product of the two velocities:

$$\gamma(V_{res}) = \gamma(V)\gamma(v_x)$$

$$\Rightarrow V_{res}^2 = V^2 + v_x^2 - \frac{Vv_x}{c^2} \qquad \text{eqn. 7}$$

Subtraction results in a speeding up of the proper time:

$$\gamma(V_{res}) = {\gamma(V)}/{\gamma(v_x)}$$

And hence:

$$\Rightarrow V_{res}^2 = \frac{V^2 - v_x^2}{1 - {v_x^2}/{c^2}} \qquad \text{eqn. 8}$$

In the frame that measures the velocity $v=\begin{bmatrix} v_x & v_y & v_z \end{bmatrix}$ time is dilated as $\gamma = \left(1 - {v^2}/{c^2}\right)^{-1/2}$ so when referring this velocity back to the original frame, although distance measurements perpendicular to the velocity V is unaffected, time reckoning is. Thus:

$$V_{res} = \sqrt{\left(V^2 + v_x^2 - \frac{Vv_x}{c^2}\right) + v_y^2\left(1 - \frac{V^2}{c^2}\right) + v_z^2\left(1 - \frac{V^2}{c^2}\right)} \qquad \text{eqn. 9}$$

Or

$$V_{res} = \sqrt{\left(\frac{V^2 - v_x^2}{1 - {v_x^2}/{c^2}}\right) + v_y^2\left(1 - \frac{V^2}{c^2}\right) + v_z^2\left(1 - \frac{V^2}{c^2}\right)} \qquad \text{eqn. 10}$$

The interferometer setup for SL signalling can test the direction field of the time dilation: the modulator can be set off in one direction with the source following at half the velocity relative to the interferometer receiver; the receiver compares a time standard against the modulator's frame and one frame will be absolutely time dilated or sped up in accordance with its true resultant absolute velocity. It is likely that such a velocity will be related or the same as the Doppler shift velocity experienced against the cosmic microwave background.

### 5.1.1. The Maximum Velocity of Closure and Separation

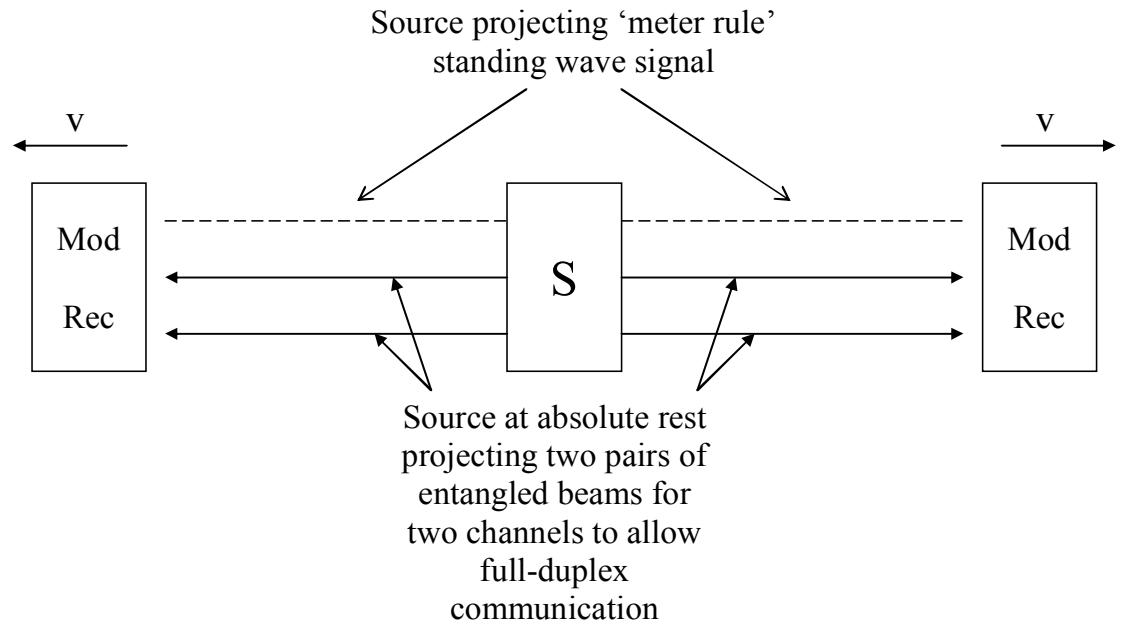

Figure 2 – The entangled source at absolute rest between two frames with interferometers

Figure 2 shows the superluminal communication setup with the source at absolute rest whilst the modulators/receivers are moving at the same velocity head-on or away from the source. The source is producing six beams, four entangled to setup a bidirectional communications channel and two beams to create a standing wave pattern with a distant reflector so that a 'meter rule' signal is created in space.

The modulator receivers can measure this meter rule and by their own time standard, carried with them, they can work out their absolute velocities in the ether, which is limited to 'c'. The communication channel lets them know each other's velocity such they can work out their maximum closure or separation rate. This is of course '2c'.

This situation, by analogy, is much the same as aircraft travelling near the speed of sound. The sound barrier would prevent them from ever exceeding the speed limit (say) in the medium and if they could only communicate by sound waves, the maximum velocity between frames would be found to be '$c_{sound}$'. "Super fast" electromagnetic signals would break this paradox and it would then be seen that the medium was preventing them from exceeding '$c_{sound}$' and that it wasn't some mysterious, logic defying concept worthy enough to be though fundamental.

## 6. The Twins Paradox in general

In the section the paradox shall be re-explored with absolute velocities. The moving twin always suffers time dilation and this can be explained *physically* by motion through an Ether not just abstractly by the interval (eqn. 3) which gives little clue to a conceptual framework to explain the effect. Also the paradox shall be seen to be a generalisation of inertial motion/free-fall maximising the proper time (eqn. 13). Figure 3 and Figure 4 show three possibilities of motion seen from the Rest Frame, $V_A$ is the starting frame which moves at constant velocity. $V_{B1}$ and $V_{B2}$ are the outward and return legs of the travelling twin's journey. On the figures, the steeper the gradient, the slower the velocity; the maximum velocity corresponds to a 45° gradient and is the null cone.

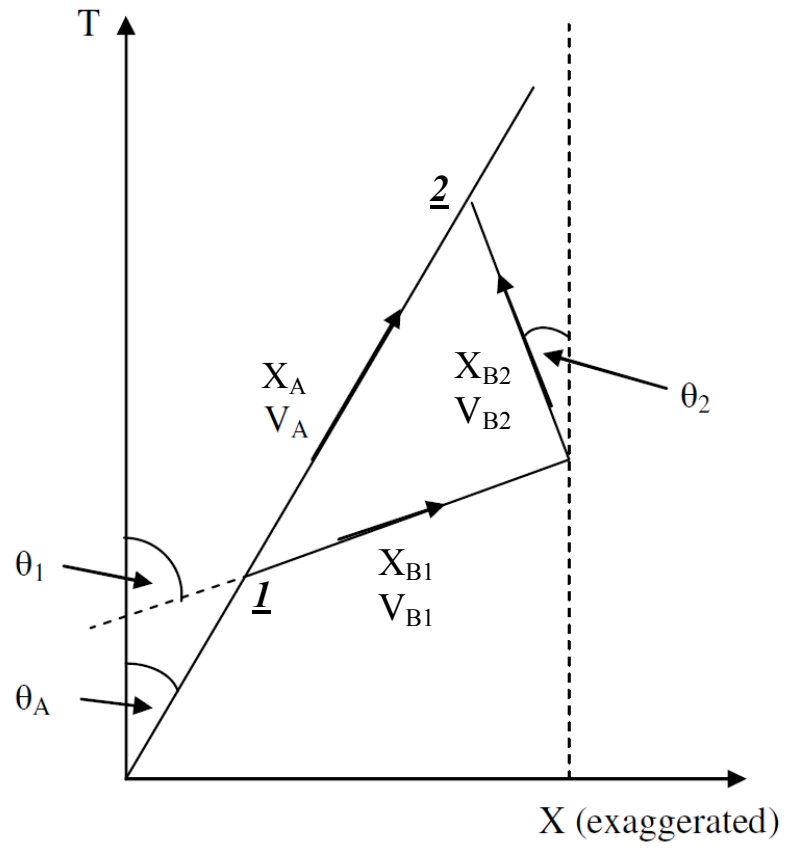

Figure 3 – Cases 1 and 2

1) The twin (frame B) always travels absolutely faster than the stay-at-home twin (frame A).

2) B moves faster than A on the first leg of the trip.

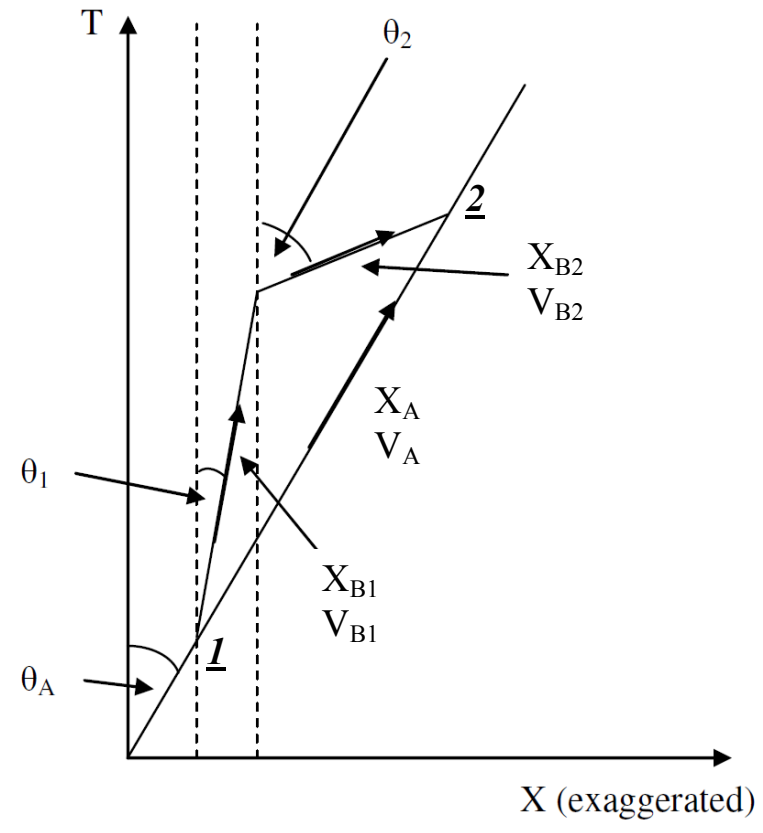

Figure 4 – Case 3

3) B moves faster than A on the second leg of the trip.

It will be seen that most time dilation occurs on the leg of the journey by frame B when it is moving faster than A which suggests that something physical is occurring relating to the way a body passes through the Ether. In the last case *frame A is time dilated* because frame B is moving slower, however most of the dilation occurs on the second leg when frame B is faster than A. The principle could be proven by some tedious trigonometry but a more general way is to calculate the proper time between events ***1*** and ***2***:

If $X = \gamma x$ and $T = \gamma t$ we prove that the proper time is maximum for frame A for event 1 to 2:

$$\int_1^2 dt_A >^? \int_1^2 dt_B \text{ if } v_A \text{ is constant}$$

Referring the measurements back to the rest frame (the time origin is set to zero):

$$\int_0^{X_A/v_A} \gamma(v_A)^{-1} dT >^? \int_0^{X_A/v_A} \gamma(v_B(T))^{-1} dT \quad \text{ineqn. 11}$$

The general proof of this could be achieved by a complicated 3D variational problem but looking at the problem, the essential and extreme elements, we are trying to prove that the proper time for frame A exceeds frame B. Thus we pursue a course to make the proper time for frame B as great as possible and note that motion along the vector $v_A$ is only relevant, as any other motion will just add time dilation. The method is to let frame B come to absolute rest for as long as is possible (no time dilation), then to complete the course at maximum speed (complete time dilation); gamma varies monotonically, so there are no turning points in the solution and this is extremal.

Thus, the time spent in frame B, at event *1*, at absolute rest is: $\frac{X_A}{v_A} - \frac{X_A}{c} = \frac{X_A}{v_A}\left(1 - \frac{v_A}{c}\right)$, where $X_A$ is the distance between events *1* and *2*. The rest of the distance is then done at maximum speed 'c' (to ensure maximum time at absolute rest) and this incurs no increase in proper time. Thus we can see that:

$$\frac{X_A}{v_A}\left(1 - \frac{v_A^2}{c^2}\right)^{\frac{1}{2}} > \frac{X_A}{v_A}\left(1 - \frac{v_A}{c}\right)$$
$$\Rightarrow \tau_A > \tau_B \ \forall v_B$$

## 7. Carrying the programme over to General Relativity

The Einstein Field Equations[14, 15, 20] were arrived at essentially heuristically and nature has chosen the simplest form. Retrospectively one would say they are "obvious": that matter gravitates was known from Newton, Special Relativity gave us mass-energy and the stress-energy tensor. Minkowski gave us space-time. On going from a flat metric, to considering rotational motion with the Equivalence Principle as a guide, Einstein considered measurements of length and time in a rotational system simulating gravity; it became apparent that space-time could be considered curved.

The mathematics of Gauss, Riemann and others provided the tools: a measure of the curvature was proportional to that which gravitates; SR taught us that it couldn't be mass alone - it had to be the stress-energy tensor. The scalar curvature was added to preserve conservation of momentum and energy ("momenergy") and the theory should reduce to Newtonian gravity in the weak limit and explain anomalies. The concept of motion itself was updated from SR's inertial frames being the simplest description of nature, to free-fall and motion with the greatest proper time being more fundamental; this was the result[14, 15, 20]:

$$R_{ij} - \frac{1}{2} g_{ij} R = \frac{8\pi G}{c^4} T_{ij} \qquad \text{eqn. 12}$$

Trajectories are computed by the geodesic equation of motion:

$$\frac{d^2 x^i}{ds^2} + \Gamma^i_{jk} \frac{dx^j}{ds} \frac{dx^k}{ds} = 0 \qquad \text{eqn. 13}$$

Most people would agree that this is one of the most (if not *the most*) profound and economic set of equations in physics – built-in is The Interval and so Lorentz Invariance and the Principle of Equivalence, ultimately describing the behaviour of massive bodies and energy constrained to a maximum velocity of 'c'. Ultimately everything macroscopic that exists and interacts locally obeys it. There can be no change to it[v]. All we ask is, how to interpret its results in the light of a phenomena not described by it – communication by entangled particles.

The EFE (eqn. 12) are involved in their solution but produce a "metric tensor" such that the interval is now represented as (covariant form):

$$ds^2 = g_{ij} dx^i dx^j \qquad \text{eqn. 14}$$

Compared to the flat space Minkowski metric, it is once again symmetrical but with off diagonal elements reflecting the infinitesimal warping of space-time over an infinitesimal distance. All information to do with time and space measure, over the region that the solution was found, is contained in it.

Real metrics are subject to constraints such as the $g_{00}$ component being positive. They generally are time varying too. A metric can be transformed to a different coordinate system by the tensor transformation:

$$g_{ij} = \frac{\partial \bar{x}^k}{\partial x^i} \frac{\partial \bar{x}^l}{\partial x^j} \bar{g}_{kl} \qquad \text{eqn. 15}$$

The barred coordinates refer to the other frame. Although complicated for the general reader, what this says is: for a region it may be possible to cancel out apparent gravitational effects by a co-ordinate transform. So a person unaware of being in an accelerating lift and believing themselves to be in a gravity field could, on standing outside the lift, realise by a coordinate transformation, that their reference frame was merely accelerating. However, a true gravity field cannot be transformed away over all space by such a transformation. A true gravity metric will not be "artificial" but the result of solving eqn. 12.

Though, in principle, general analytical solution of the EFE is complicated[vi] solutions exist which will be sufficient for our purposes in the following arguments. Some will represent real gravity fields, some a rotating coordinate system.

## 7.1. Clock synchronisation

In SR one can theoretically move clocks infinitely slowly in one frame until they have time-like separation. However due to the failure in the relativity of simultaneity, another frame will not view these clocks in synchrony. The Ether transform (section 5) and superluminal signals permit clock synchronisation in all frames. Section 5.1.1 gave an example of the interferometer operating between frames.

In GR the clock synchronisation problem is more difficult, metrics can be time varying. The relation between proper time at a point to the global coordinate system[vii] used is given by:

$$d\tau = \frac{1}{c} \sqrt{g_{00}} dx^0 \qquad \text{eqn. 16}$$

---

[v] Apart from a "Cosmological Constant" on the RHS.

[vi] *And* the geodesic equation too, to then re-compute the stress-energy tensor from the rearrangement of matter.

[vii] A system we *choose* to use, it is arbitrary.

The difference in time (measured in the global coordinate system) between two infinitesimally separated points[viii] is:

$$\Delta x^0 = -\frac{g_{0\alpha}}{g_{00}} dx^\alpha \qquad \text{eqn. 17}$$

Where α is a space co-ordinate. Its derivation is based on light speed signals (section 7.2).

If a metric can be written (or transformed by eqn. 15) to a coordinate system where all the components are independent of the time coordinate $x^0$ over all of space, then gravitational field is called static. Specifically, the components $g_{0\alpha}$ are zero. In this case all clocks can be synchronised and $x^0$ is called the "World Time".

To prove general synchronisation of clocks by light speed signals around a closed path, the contour integration is performed:

$$\Delta x^0 = -\oint \frac{g_{0\alpha}}{g_{00}} dx^\alpha \qquad \text{eqn. 18}$$

If this is zero then we can synchronise clocks. Most generally, it is not, as a simple rotating metric proves[ix]:

$$ds^2 = \left(c^2 - \Omega^2 r^2\right)dt^2 - 2\Omega r^2 dt d\theta - dz^2 - r^2 d\theta^2 - dr^2 \qquad \text{eqn. 19}$$

7.1.1. Global Synchronisation by use of Superluminal Signals

Figure 2 showed the interferometer setup between two frames in relative motion. It is possible to "beam in" world time and a time standard far from where the gravitational field is minimum. In principle the path lengths are adjusted (section 7.2) and gravitational time dilation (eqn. 16) can be accounted for, such that receiver is the right distance from the source and modulator to be receiving SL signals.

The assumption in eqn. 17 and eqn. 20 is that local reckoning of light transit time (from the metric) is used to compute the global time, if it can be computed at all. In our method, if a receiver can lock into a beam from the time standard in flat space, such that it is the correct distance from the source and modulator[x], then all clocks can be synchronised to a World Time in any coordinate system. This is because no components are used from the metric (apart from setting the path lengths) and by definition SL communication is not dependent on the distance between points.

7.2. Distance measurement

Stated without proof[14] the interval between the departure and arrival of a signal at a point between another infinitesimally close point is given by:

$$dx^{0(\text{event 1})} = \frac{1}{g_{00}}\left(-g_{0\alpha}dx^\alpha - \sqrt{\left(g_{0\alpha}g_{0\beta} - g_{\alpha\beta}g_{00}\right)dx^\alpha dx^\beta}\right) \qquad \text{eqn. 20}$$

$$dx^{0(\text{event 2})} = \frac{1}{g_{00}}\left(-g_{0\alpha}dx^\alpha + \sqrt{\left(g_{0\alpha}g_{0\beta} - g_{\alpha\beta}g_{00}\right)dx^\alpha dx^\beta}\right)$$

The amount of proper time that has elapsed in the measurement is computed by multiplying the difference in this interval by $\sqrt{g_{00}}/c$ according to eqn. 16. Multiplying again by c/2 (two leg trip) gives the relation to infinitesimal distance measured locally at some point in the gravity field to the global coordinates:

$$dl^2 = \left(-g_{\alpha\beta} + \frac{g_{0\alpha}g_{\alpha\beta}}{g_{00}}\right)dx^\alpha dx^\beta \qquad \text{eqn. 21}$$

Metrics are generally time dependent and it becomes meaningless to try and integrate dl[xi] – thus in GR the distance between bodies cannot be found in most cases.

7.2.1. Global Distance measurement by use of Superluminal Signals

Once again, a time standard far away from the gravitational field in flat space, can communicate what a standard length is by time signal; it would send the time it took a light signal to travel a unit distance. Now although the space in between the modulator, source and receiver may be sifting, provided that the receiver could lock on to a beam and receive intelligible data[xii], it would pick up that distance standard. Thus a "World length" would be communicated to any position in the gravity field.

In principle, enough sources and receivers placed throughout the space in a network would give sufficient resolution to map distance to any location, in absolute units – though it would be distance computed at one instant of universal world time.

viii $x^0 + \Delta x^0 = x^0 + \frac{1}{2}\left(dx^{0(\text{event 2})} + dx^{0(\text{event 1})}\right)$ dx from eqn. 20

ix $g_{tt} = \left(c^2 - \Omega^2 r^2\right)$ and $g_{t\theta} = -2\Omega r^2 d\theta dt$

x In general the source will not be equidistant between modulator and receiver.

xi The integral would depend on the world line between two given points.

xii It needs to be at the equivalent distance the modulator is from the source.

## 7.3. Referring it all back to the Universal Rest Frame

Now, just to illustrate concepts, consider a Lorentz boosted metric (by a coordinate transform eqn. 15). The boost renders the gravitating source moving so that in the global coordinates of the problem, far from the source can be considered at rest in the Universal Rest Frame. For instance, a boosted Schwarzchild metric[21] in isotropic coordinates is developed as follows (in Geometrised units):

$$ds^2 = -\left(\frac{1-M/2r}{1+M/2r}\right)^2 dt^2 + \left(1+\frac{M}{2r}\right)\left(dx^2+dy^2+dz^2\right) \qquad \text{eqn. 22}$$

where $r^2 = x^2+y^2+z^2$

and $M$ is the mass of the gravitating body

A boost is applied:

$$\bar{t} = \left(1-v^2\right)^{-1/2}\left(t+vx\right)$$
$$\bar{x} = \left(1-v^2\right)^{-1/2}\left(x+vt\right)$$
$$\bar{y} = y$$
$$\bar{z} = z$$

And the metric takes on the form:

$$ds^2 = \left(1+A\right)^4\left(-d\bar{t}^2 + d\bar{x}^2 + d\bar{y}^2 + d\bar{z}^2\right) + \left[\left(\frac{1-A}{1+A}\right)^2 - \left(1+A\right)^4\right]\frac{\left(d\bar{t}-vd\bar{x}\right)^2}{\left(1-v^2\right)} \qquad \text{eqn. 23}$$

Where

$$A = \frac{M}{2r} = \frac{\mu\left(1-v^2\right)}{2\left[\left(\bar{x}-v\bar{t}\right)^2 + \left(1-v^2\right)\left(\bar{y}^2+\bar{z}^2\right)\right]^{1/2}}$$

and

$$\mu = M/\sqrt{\left(1-v^2\right)}$$

Obviously this is complicated but we can track the coefficients enough in front of the coordinates to compute the proper time and "proper length" at an instant of Global Universal Time via eqn. 16 and eqn. 21. Locally space is perceived as flat so we then multiply by the Ether transform, for a frame moving at velocity V[xiii], to arrive at the differential Universal Ether Transform at a point in the space and instant in time (eqn. 24).

The transform can be integrated as a function of universal time, T, over all space to find distance between all bodies *at an instant in time*. The fact that the distance could be space-like beyond local light cones may have little point, for us, in today's world. Mass-energy is constrained to move at or below 'c', as far as we know. To make any use of such distances, one would have to "warp" through space – it would then be regarded as a cosmic map.

## 8. Conclusion and Further work

It is hoped that the reader will agree with development of an Ether model in this paper, such that objective time and distance standards can be communicated to all space, from a position in space at absolute rest and devoid of gravitational field.

We are not saying that the picture the Ether Transform presents is graceful or even a preferable system to do analysis in, it doesn't form a group and so transformation is not easy compared to the elegant mathematics of tensors and space-time. It may turn out that a return to the Ether concept will spur new developments and give it mechanical properties[17] such that novel forms of propulsion will result. There is more to the vacuum, we believe.

$$\begin{bmatrix} dT \\ dX \\ dY \\ dZ \end{bmatrix} = \begin{bmatrix} \frac{c}{\sqrt{g(T)_{00}}} & 0 & 0 & 0 \\ 0 & \frac{1}{\sqrt{-g(T)_{XX} + \frac{g(T)_{0X}\,g(T)_{XX}}{g(T)_{00}}}} & \frac{1}{\sqrt{-g(T)_{XY} + \frac{g(T)_{0X}\,g(T)_{XY}}{g(T)_{00}}}} & \frac{1}{\sqrt{-g(T)_{XZ} + \frac{g(T)_{0X}\,g(T)_{XZ}}{g(T)_{00}}}} \\ 0 & \frac{1}{\sqrt{-g(T)_{YX} + \frac{g(T)_{0Y}\,g(T)_{YX}}{g(T)_{00}}}} & \frac{1}{\sqrt{-g(T)_{YY} + \frac{g(T)_{0Y}\,g(T)_{YY}}{g(T)_{00}}}} & \frac{1}{\sqrt{-g(T)_{YZ} + \frac{g(T)_{0Y}\,g(T)_{YZ}}{g(T)_{00}}}} \\ 0 & \frac{1}{\sqrt{-g(T)_{ZX} + \frac{g(T)_{0Z}\,g(T)_{ZX}}{g(T)_{00}}}} & \frac{1}{\sqrt{-g(T)_{ZY} + \frac{g(T)_{0Z}\,g(T)_{ZY}}{g(T)_{00}}}} & \frac{1}{\sqrt{-g(T)_{ZZ} + \frac{g(T)_{0Z}\,g(T)_{ZZ}}{g(T)_{00}}}} \end{bmatrix}$$

$$\times \begin{bmatrix} \gamma(V) & 0 & 0 & 0 \\ 0 & 1+(\gamma-1)\frac{V_x^2}{V^2} & (\gamma-1)\frac{V_xV_y}{V^2} & (\gamma-1)\frac{V_xV_z}{V^2} \\ 0 & (\gamma-1)\frac{V_xV_y}{V^2} & 1+(\gamma-1)\frac{V_y^2}{V^2} & (\gamma-1)\frac{V_yV_z}{V^2} \\ 0 & (\gamma-1)\frac{V_xV_z}{V^2} & (\gamma-1)\frac{V_yV_z}{V^2} & 1+(\gamma-1)\frac{V_z^2}{V^2} \end{bmatrix} \begin{bmatrix} dt \\ dx \\ dy \\ dz \end{bmatrix} \qquad \text{eqn. 24}$$

---

[xiii] This V is reckoned at the locally infinitesimally flat space point in question and can be referred back to distant flat space by the metric components; this would then directly relate the $dX_i$ to $dx_i$ by one equivalent Ether Transform i.e. one matrix.

Appendix 1 Against the No-signalling Theorem

The "No-signalling Theorem"[2-4] contains an omission or restriction in logic – an averaged or flat expectation value is used so that spatial variation in expectation is not included. The prediction it would make for attempting to send information by entangled photons is:

$$\rho_2(\text{bit } 0) = (1/2)|\psi_V\rangle\langle\psi_V| + (1/2)|\psi_H\rangle\langle\psi_H|$$
$$\rho_2(\text{bit } 1) = (1/2)|\psi_V\rangle\langle\psi_V| + (1/2)|\psi_H\rangle\langle\psi_H|$$

Which really amounts to saying that photons or mass-energy was used to affect the transmission. We agree as we believe in a *physical* Universe too, as opposed to a magical one.

In our particular case bit 1 can be signalled by performing a measurement on the entangled state: $\frac{1}{\sqrt{2}}\left(|H\rangle\otimes|V\rangle+|V\rangle\otimes|H\rangle\right)$. Done at either end, both ends would end in a mixed state. Our wave-function is represented thus:

$$\psi_{AB} = \begin{pmatrix} |H_A\rangle|V_B\rangle/\sqrt{2} & 0 \\ 0 & 0 \end{pmatrix} \text{ or } \begin{pmatrix} 0 & 0 \\ 0 & |V_A\rangle|H_B\rangle/\sqrt{2} \end{pmatrix}$$

And the reduced density operator at the detector after the interferometer with no interference from the mixed state is:

$$\rho_B(\text{bit } 1) = tr_A\left[\begin{pmatrix} |H_A\rangle|V_B\rangle/\sqrt{2} & 0 \\ 0 & 0 \end{pmatrix}\begin{pmatrix} \langle H_A|\langle V_B|/\sqrt{2} & 0 \\ 0 & 0 \end{pmatrix}\right] + tr_A\left[\begin{pmatrix} 0 & 0 \\ 0 & |V_A\rangle|H_B\rangle/\sqrt{2} \end{pmatrix}\begin{pmatrix} 0 & 0 \\ 0 & \langle V_A|\langle H_B|/\sqrt{2} \end{pmatrix}\right]$$

$$= (1/2)|V_B\rangle\langle V_B| + (1/2)|H_B\rangle\langle H_B|$$

The absence of the modulator and hence the interference possible from the superposition state gives a different expectation:

$$\psi_{AB} = \begin{pmatrix} |H_A\rangle|V_B\rangle/\sqrt{2} & 0 \\ 0 & |V_A\rangle|H_B\rangle/\sqrt{2} \end{pmatrix}$$

This wave-function changes on passing through the interferometer (path length difference $e^{ikx}$) and the Faraday rotators:

$$\psi'_{AB} = R_{V\left(-\frac{\pi}{4}\right)} \otimes \begin{pmatrix} |H_A\rangle|e^{ikx}V_B\rangle/\sqrt{2} & 0 \\ 0 & 0 \end{pmatrix} + R_{H\left(\frac{\pi}{4}\right)} \otimes \begin{pmatrix} 0 & 0 \\ 0 & |V_A\rangle|H_B\rangle/\sqrt{2} \end{pmatrix}$$

$$= \begin{pmatrix} \frac{1}{\sqrt{2}} & \frac{1}{\sqrt{2}} \\ -\frac{1}{\sqrt{2}} & \frac{1}{\sqrt{2}} \end{pmatrix}\begin{pmatrix} |H_A\rangle|e^{ikx}V_B\rangle/\sqrt{2} & 0 \\ 0 & 0 \end{pmatrix} + \begin{pmatrix} \frac{1}{\sqrt{2}} & -\frac{1}{\sqrt{2}} \\ \frac{1}{\sqrt{2}} & \frac{1}{\sqrt{2}} \end{pmatrix}\begin{pmatrix} 0 & 0 \\ 0 & |V_A\rangle|H_B\rangle/\sqrt{2} \end{pmatrix}$$

$$= \begin{pmatrix} |H_A\rangle|e^{ikx}V_B\rangle/2 & -|V_A\rangle|H_B\rangle/2 \\ -|H_A\rangle|e^{ikx}V_B\rangle/2 & |V_A\rangle|H_B\rangle/2 \end{pmatrix}$$

On change of basis this is recognised as ('D' for diagonal):

$$\psi'_{AB} = \begin{pmatrix} |D_A\rangle|e^{ikx}D_B\rangle/\sqrt{2} & 0 \\ 0 & |D_A\rangle|D_B\rangle/\sqrt{2} \end{pmatrix}$$

Forming the density matrix $\rho_{AB} = |\psi'_{AB}\rangle\langle\psi'_{AB}|$ and tracing out system A, the photons exiting the interferometer impinge on the detector giving the reduced density operator:

$$\rho_B(\text{bit } 0) = \frac{\left(e^{ikx}+1\right)}{2}|\psi_D\rangle\langle\psi_D|$$

Clearly the path length has provided an interference term. The photon hasn't disappeared at the null-point! Moving the detector along the expectation becomes the same as bit 1. The correct interpretation of the analysis[2-4] is then, "a photon/mass-energy was used for transmission"; this is hardly profound.

Appendix 2 Space-time view of Twin Paradox

A classic twin's paradox is the case of a travelling twin moving at 0.6c for 20 years (so a distance of 12 light-years) but the moving twin only experiences 16 years. This is explained conventionally by the invariant interval relating what the two twins calculate in their system of measurement:

$$T^2 - D^2 = t^2 - d^2$$
$$20^2 - 12^2 = t^2 - 0$$

So 16 years.

Figure 5 below shows a space-time diagram of the home's perspective of beacons sent from home to the moving twin. The moving twin is time dilated and it counts 16 beacons.

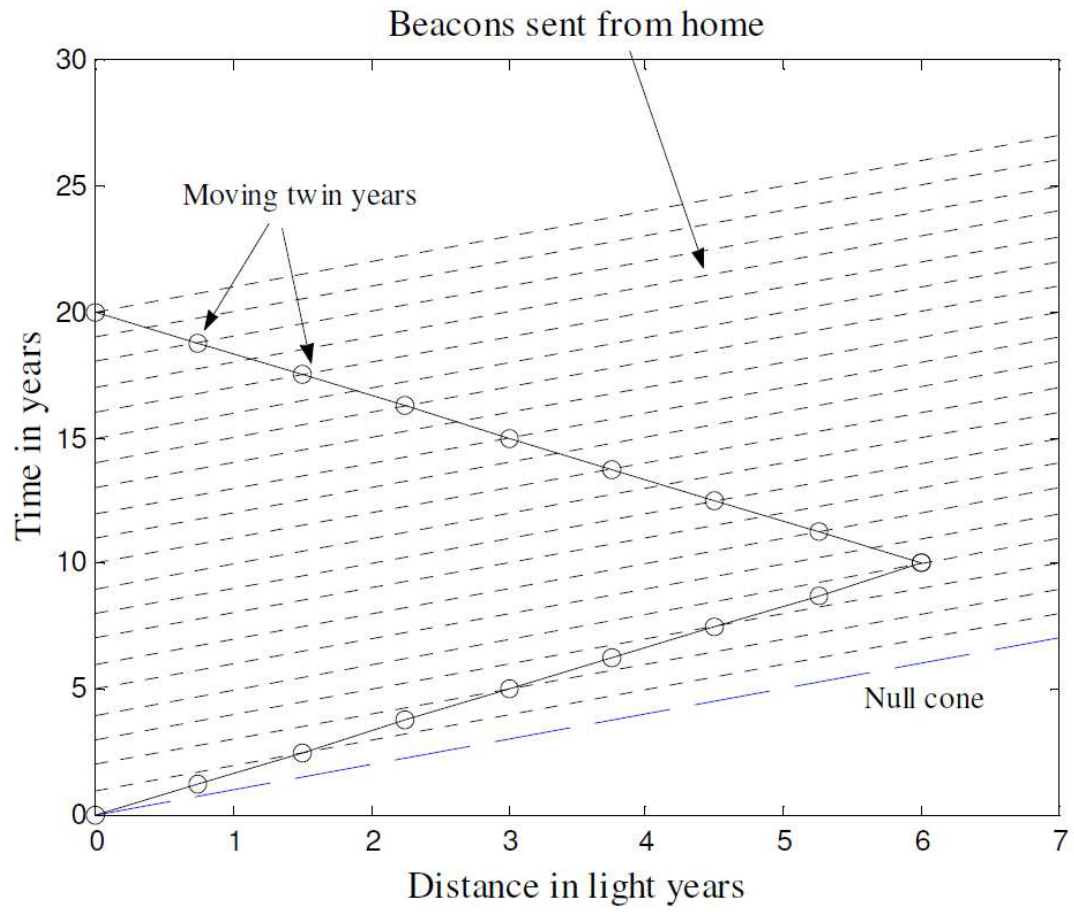

Figure 5 – Beacons sent from home

Figure 6 below shows a space-time diagram of the home's perspective of beacons sent from the moving to twin home. Once again, the moving twin is time dilated but the home twin still counts 16 beacons.

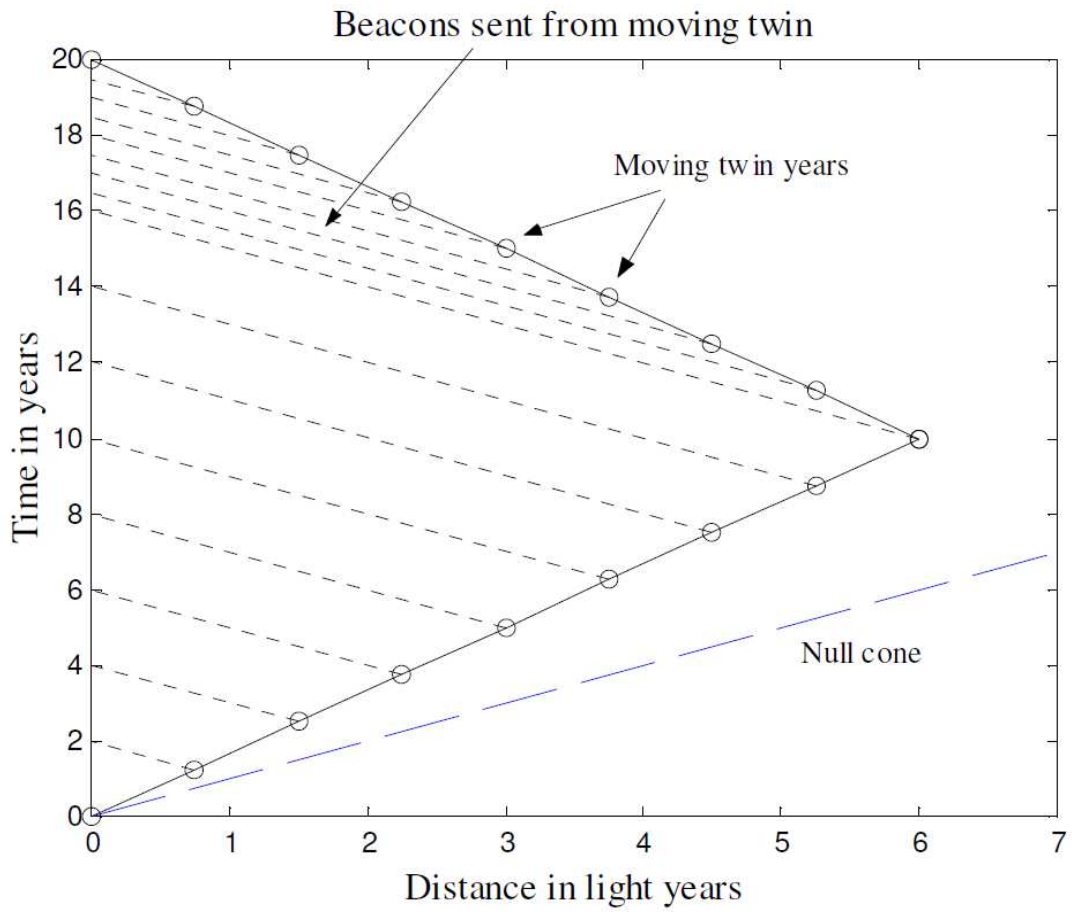

Figure 6 – Beacons sent from moving twin

Appendix 3 Analysis of world lines in Doppler shift between two frames

We shall look at the Doppler shift from the perspective of the frame regarded as stationary (frame A) in a simple 1st order analysis. First consider beacons sent from frame B to frame A.

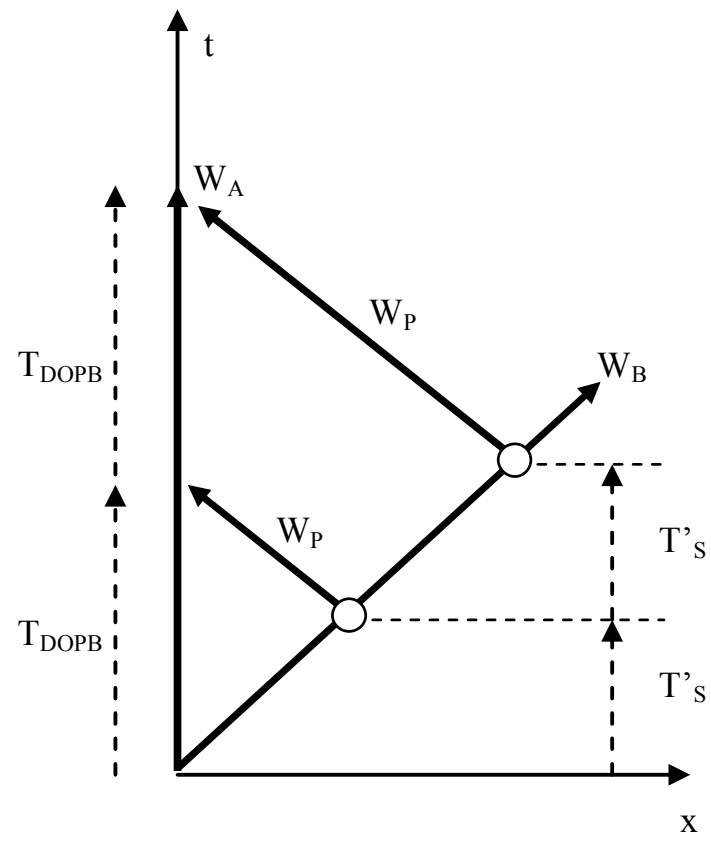

Figure 7 – Beacons from B to A

Figure 7 shows the world lines (in frame A's perspective) of frame A ($W_A$), frame B ($W_B$) and photons ($W_P$) emitted by twin B carrying a time standard $T_S$ which we say is time dilated by some factor $\Gamma_B(v)$ where v is the velocity. Writing world lines as:

$$W_B : t_B = {x_B}/{v}$$

$$W_P : t_P = {-x_P}/{c} + T_{DOPB} \text{ (for 1st beacon)}$$

At intersection of world lines $t_B = t_P = T'_S$

Thus if $T'_S = {T_S}/{\Gamma_B(v)}$ and $F_S = T_S^{-1}$ then:

$$F_{DOPB} = \frac{F_S \Gamma_B(v)}{1 + {v}/{c}} \qquad \text{eqn. 25}$$

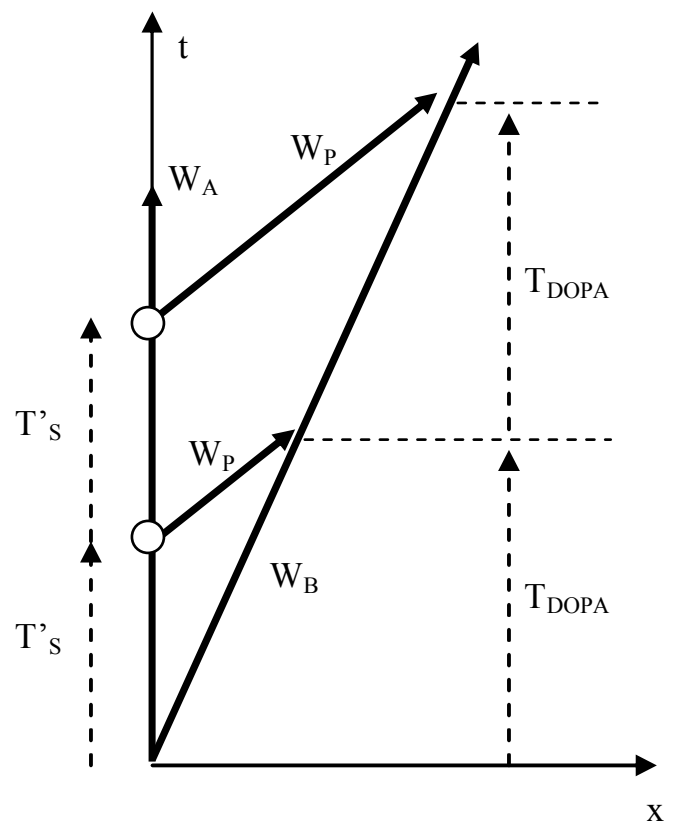

Figure 8 – Beacons from A to B

Figure 8 shows the reverse situation, this time the time standard is dilated by some factor $\Gamma_A(v)$:

$$W_B : t_B = {x_B}/{v}$$

$$W_P : t_P = {x_P}/{c} + T'_S \text{ (for 1st beacon)}$$

At intersection of world lines $t_B = t_P = T_{DOPA}$

Thus if $T'_S = {T_S}/{\Gamma_A(v)}$ and $F_S = T_S^{-1}$ then:

$$F_{DOPB} = F_S \Gamma_A(v)\left(1 - {v}/{c}\right) \qquad \text{eqn. 26}$$

By *reciprocity* of the Doppler shift if:

$$\Gamma_A = \frac{1}{\sqrt{1 - {v^2}/{c^2}}} \text{ and } \Gamma_B = \sqrt{1 - {v^2}/{c^2}}$$

The familiar Doppler shift equations result:

$$F_{DOP} = \frac{F_S}{\sqrt{1 - {v^2}/{c^2}}}\left(1 + {v}/{c}\right) \qquad \text{eqn. 27}$$